\documentclass{article}

\PassOptionsToPackage{numbers, compress}{natbib}
\usepackage[preprint]{neurips_2026}
\usepackage[utf8]{inputenc}
\usepackage[T1]{fontenc}
\usepackage{hyperref}
\usepackage{url}
\usepackage{booktabs}
\usepackage{graphicx}
\usepackage{amsfonts}
\usepackage{amsmath}
\usepackage{amssymb}
\usepackage{nicefrac}
\usepackage{microtype}
\usepackage{xcolor}
\usepackage{algorithm}
\usepackage{algpseudocode}
\usepackage{pifont}
\usepackage{tabularx}
\usepackage{array}
\usepackage{amsthm}
\newcommand{\pmark}{\(\triangle\)}
\newcommand{\cmark}{\textcolor{green!60!black}{\ding{51}}}
\newcommand{\xmark}{\textcolor{red!70!black}{\ding{55}}}

\title{Safe Multi-Agent Behavior Must Be Maintained, Not Merely Asserted: Constraint Drift in LLM-Based Multi-Agent Systems}

\author{%
\begin{tabular}{c}
{\normalsize\bfseries Tianxiao Li\textsuperscript{1} \quad
Yixing Ma\textsuperscript{2} \quad
Haiquan Wen\textsuperscript{1} \quad
Zhenglin Huang\textsuperscript{1}} \\[-0.1em]
{\normalsize\bfseries Qianyu Zhou\textsuperscript{4} \quad
Zeyu Fu\textsuperscript{3} \quad
Guangliang Cheng\textsuperscript{1,*}} \\[0.7em]
{\normalsize \textsuperscript{1}University of Liverpool, UK} \\
{\normalsize \textsuperscript{2}University of Nottingham, UK} \\
{\normalsize \textsuperscript{3}University of Exeter, UK} \\
{\normalsize \textsuperscript{4}University of Tokyo, Japan} \\[0.3em]
{\normalsize \textsuperscript{*}Corresponding author: \texttt{guangliang.cheng@liverpool.ac.uk}}
\end{tabular}
}

\begin{document}

\maketitle

\begin{abstract}
Modern LLM based agents are no longer passive text generators. They read repositories, call tools, browse the web, execute code, maintain memory, communicate with other agents, and act through long horizon workflows. This shift moves the unit of safety. A system may produce a compliant final answer while leaking private information through an internal message, delegating authority beyond its original scope, calling an external tool with sensitive context, or losing the evidence needed to reconstruct why an action was allowed. We argue that many emerging failures in LLM-based multi-agent systems share a common structure: safety critical constraints do not remain operative throughout the trajectory. We call this phenomenon \emph{constraint drift}: the loss, distortion, weakening, or relaxation of constraints as they pass through memory, delegation, communication, tool use, audit, and optimization. The position taken here is that safe multi-agent behavior must be maintained, not merely asserted. Prompts, guardrails, tool schemas, access control, and final output checks are necessary, but they are insufficient unless constraints remain fresh, inherited, enforceable, and auditable across execution. We propose \emph{Constraint State Governance} as a research paradigm for LLM-based multi-agent systems. In this paradigm, safety-critical constraints are maintained as explicit execution state, while constraint-native reinforcement learning improves utility only within maintained safety boundaries. The goal is not to freeze agentic systems under rigid rules, but to make safety operational across the trajectories through which modern agents actually act.
\end{abstract}

\section{Introduction}

Modern LLM-based agents are becoming execution systems instead of text generators. Coding agents make this very visible: systems such as Codex, Claude Code, SWE-agent, and OpenHands can inspect repositories, edit files, run commands, execute tests, and operate inside software environments \citep{openai2025codex, anthropic2026claudecode, jimenez2024swebenchlanguagemodelsresolve, yang2024sweagentagentcomputerinterfacesenable, wang2025openhandsopenplatformai}. This paper focuses on LLM-based multi-agent systems that act through tools, memory, delegation, protocols, and executable actions.

This shift moves the unit of safety. When the setting is a single response, safety is often judged from the answer shown to the user. In a multi-agent execution setting, the final answer is only the visible endpoint of a longer trajectory \citep{hammond2025multiagentrisksadvancedai}. Consider a coding assistant working on a private repository under constraints: do not expose API keys, do not send proprietary code to external services, do not modify production configuration, do not bypass tests, and do not delete files outside the relevant module. The final patch may compile and the final message may look compliant, while the trajectory has already leaked a token in an internal message, sent code through a tool call, expanded a narrow permission into broader edit authority, or deleted a failing test during cleanup. The safety failure lies not only in what the system says, but in how it acted.

Recent work suggests this surface is already exposed. AgentLeak shows that privacy leakage often happens through inter-agent messages, shared memory, and tool arguments that output only audits miss \citep{yagoubi2026agentleakfullstackbenchmarkprivacy}. PAC-Bench studies privacy constrained collaboration; Prompt Infection shows malicious instructions propagating across agents; AgentDojo, InjecAgent, and ToolEmu expose prompt injection and tool use failures; and Colosseum argues for auditing communication and action traces \citep{park2026pacbenchevaluatingmultiagentcollaboration, lee2024promptinfectionllmtollmprompt, debenedetti2024agentdojodynamicenvironmentevaluate, zhan2024injecagentbenchmarkingindirectprompt, ruan2024identifyingriskslmagents, nakamura2026colosseumauditingcollusioncooperative}. Together, these results point to the same problem: the risk is often inside the trajectory, before the final output is produced.

We call this phenomenon \emph{constraint drift}: the loss of operational force of a safety constraint across an LLM-based multi-agent trajectory. We use five diagnostic modes: memory drift, authority drift, information-flow drift, accountability drift, and utility-induced drift, marking where a constraint stops being fresh, inherited, enforceable, or auditable. The list is not meant to be a closed taxonomy. Unlike reward hacking, specification gaming, or goal misgeneralization, which concern flawed objectives, proxy exploitation, or unintended goals \citep{amodei2016concreteproblemsaisafety, langosco2023goalmisgeneralizationdeepreinforcement, bondarenko2025demonstratingspecificationgamingreasoning}, constraint drift names a preservation failure inside agentic execution. Its LLM-specific challenge is that constraints are natural language objects that can be summarized, reinterpreted, routed through tools, and justified after the fact.

\textbf{The position of this paper is: safe multi-agent behavior must be maintained, not merely asserted.} Output level safety falls short for agentic systems. A final answer can read as compliant even when the trajectory itself has already broken privacy, authority, audit, or file boundary constraints. Prompts, guardrails, tool schemas, access control, filters, and logs are only useful when they keep constraints tied to the actions those constraints are meant to govern. Agent safety benchmarks therefore need to look at whether constraints are preserved across the trajectory. Looking only at the final output is not enough.

This paper makes two contributions. First, it introduces constraint drift as an analytical framework for trajectory level safety failures in LLM-based multi-agent systems. Second, it proposes \emph{Constraint State Governance} as a design paradigm in which safety critical constraints are represented as signed state, inherited through scoped delegation, checked before critical actions, recorded for audit, and supplied to learning. We further show how \emph{Constraint Native Reinforcement Learning} can optimize utility only within admissible trajectories, rather than treating mandatory constraints as soft penalties. An empirical replay study on AgentLeak grounds the argument: output only filtering can make final answers look safe while internal channels remain exposed, whereas transition level governance makes those violations visible, auditable, and catchable.

The goal is not to freeze LLM-based agents under rigid rules or to claim that every human norm can be perfectly formalized. The claim is narrower: when a constraint is safety critical for a trajectory, the system should not rely only on prompt repetition, local memory, final output inspection, or post hoc explanation. It should make the constraint harder to lose.

\begin{figure*}[!htbp]
    \centering
    \includegraphics[width=0.98\textwidth]{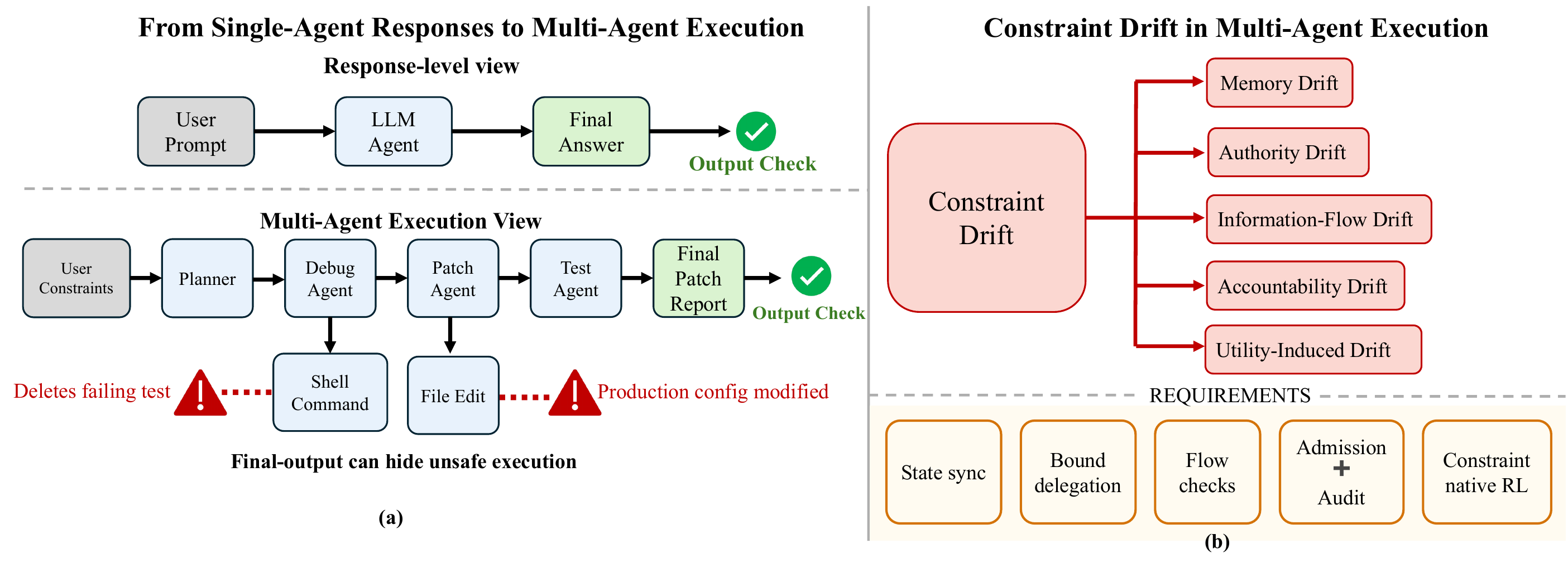}
    \caption{(a) Multi-agent execution shifts safety from final-answer checking to trajectory-level constraint preservation; (b) constraint drift identifies where constraints lose operational force across memory, authority, information flow, accountability, and optimization.}
    \label{fig:constraint_drift_reframing}
\end{figure*}

\section{Constraint Drift as a Systems-Level Preservation Failure}

Constraint drift names a preservation failure. It does not cover every unsafe behavior an agent might produce. An LLM-based multi-agent system may begin with explicit safety critical constraints: which files may be edited, which commands may be executed, which secrets may leave the repository, which agents may hold write authority, which tests must remain intact, and which actions require audit evidence. The central question is not whether these constraints were stated at the start. The question is whether they remain operative once the system summarizes context, delegates work, calls tools, edits files, runs commands, updates memory, and optimizes for task success.

We define \emph{constraint drift} as the loss of operational force of safety critical constraints across an LLM-based multi-agent trajectory. A constraint has operational force when it is available at the point of action, attached to the authority under which the action is proposed, checkable against the action's semantic effect, and reconstructable after execution\citep{amodei2016concreteproblemsaisafety, bondarenko2025demonstratingspecificationgamingreasoning}. Drift occurs when one of these properties weakens: a no disclosure rule is omitted from a tool call context, read access becomes de facto write authority, a failing test is deleted during cleanup, or a destructive command is logged without evidence that it was authorized. This is a systems level preservation failure: each local agent may appear compliant, while the constraint still appears in the prompt, summary, or final explanation but no longer governs the action that matters. This distinguishes constraint drift from reward hacking, specification gaming, goal misgeneralization, and prompt injection, which concern flawed objectives, benchmark loopholes, unintended learned goals, or adversarial context manipulation \citep{ langosco2023goalmisgeneralizationdeepreinforcement, lee2024promptinfectionllmtollmprompt, debenedetti2024agentdojodynamicenvironmentevaluate, zhan2024injecagentbenchmarkingindirectprompt}. It is also narrower than broad multi-agent risk categories such as miscoordination, conflict, collusion, information asymmetry, and emergent agency \citep{hammond2025multiagentrisksadvancedai, dafoe2020openproblemscooperativeai, motwani2025secretcollusionaiagents}. These failures can interact, but constraint drift isolates one mechanism: safety critical conditions stop functioning as operative constraints while agents coordinate, communicate, delegate, use tools, and optimize.

\subsection{Drift Modes from Trajectory Interfaces}

The five modes in Table~\ref{tab:drift_modes} are diagnostic rather than mutually exclusive. They are derived from where LLM-based multi-agent trajectories repeatedly transfer constraints: state and memory, delegation and authority, communication and tool I/O, execution and audit, and optimization. In coding agents, these interfaces appear as repository inspection, context summarization, file edits, shell commands, test execution, cleanup, memory updates, and review. The modes can overlap: memory drift may cause information flow drift; authority drift may surface as an audit gap; utility induced drift may amplify other modes by selecting trajectories that look successful because constraints have weakened. The point is diagnostic. When an agent leaks a token, deletes a test, violates a file boundary, or modifies production configuration, constraint drift asks where the relevant constraint stopped governing behavior. Final-output safety cannot identify which of these interfaces failed. However, a trajectory-level view can.

\begin{table}[!htbp]
\centering
\scriptsize
\setlength{\tabcolsep}{3.5pt}
\renewcommand{\arraystretch}{1.12}
\caption{Diagnostic drift modes derived from recurring interfaces in LLM-based multi-agent trajectories.}
\label{tab:drift_modes}
\begin{tabular}{p{0.21\linewidth} p{0.19\linewidth} p{0.50\linewidth}}
\toprule
\textbf{Interface} & \textbf{Mode} & \textbf{Coding agent failure} \\
\midrule
State and memory &
Memory drift &
Constraint is omitted or retrieved imprecisely; ``do not delete outside auth'' becomes ``clean up irrelevant files'' \citep{liu2023lostmiddlelanguagemodels, laban2025llmslostmultiturnconversation, du2026memoryautonomousllmagentsmechanisms}. \\

Delegation and authority &
Authority drift &
Delegated scope widens; read only inspection becomes permission to edit production config \citep{south2025authenticateddelegationauthorizedai, prakash2026aipagentidentityprotocol, https://doi.org/10.5281/zenodo.19672575}. \\

Communication and tool I/O &
Information flow drift &
Sensitive content crosses messages, memory, tool arguments, or external queries; an API key enters an internal message \citep{yagoubi2026agentleakfullstackbenchmarkprivacy, 10.1145/2619091}. \\

Execution and audit &
Accountability drift &
Action is logged without reconstructable state, authority, or evidence; a file deletion cannot be justified as authorized \citep{nakamura2026colosseumauditingcollusioncooperative, motwani2025secretcollusionaiagents, zhou2026governingdynamiccapabilitiescryptographic}. \\

Optimization and adaptation &
Utility induced drift &
Task success improves by weakening constraints treated as soft costs; a failing test is deleted before reporting success \citep{achiam2017constrainedpolicyoptimization, tessler2018rewardconstrainedpolicyoptimization, kushwaha2026surveysafereinforcementlearning, lightman2023letsverifystepstep}. \\
\bottomrule
\end{tabular}
\end{table}

\section{Why Existing Mechanisms Are Necessary but Insufficient}

Existing agent safety mechanisms are necessary, but they usually control local units of safety: a final response, a tool call, a workflow node, a delegation token, or a reward signal. Response safeguards reduce unsafe text; tool and action benchmarks expose risky tool use, prompt injection, browser actions, file operations, shell commands, and code execution failures \citep{debenedetti2024agentdojodynamicenvironmentevaluate, zhan2024injecagentbenchmarkingindirectprompt, ruan2024identifyingriskslmagents, evtimov2025waspbenchmarkingwebagent}. Agent frameworks and protocols make systems composable by routing tasks, sharing context, calling tools, and coordinating agents \citep{openai_agents_sdk, anthropic_mcp, google_a2a, langgraph, wu2023autogenenablingnextgenllm}. Multi-agent benchmarks reveal leakage, privacy failure, and collusion through internal messages, shared memory, tool arguments, and communication traces \citep{yagoubi2026agentleakfullstackbenchmarkprivacy, park2026pacbenchevaluatingmultiagentcollaboration, nakamura2026colosseumauditingcollusioncooperative}. Delegation, capability, safe RL, and process supervision address important pieces by binding authority or optimizing under constraints \citep{prakash2026aipagentidentityprotocol, https://doi.org/10.5281/zenodo.19672575, achiam2017constrainedpolicyoptimization}. The gap is compositional: these mechanisms do not by themselves keep a safety critical constraint inherited, fresh, enforceable, and auditable as a task moves across agents, memory, tools, file edits, shell commands, and learning signals. A locally valid action can still be semantically unsafe, and a constraint can drift before optimization can respect it. The missing layer is therefore not another isolated guardrail, benchmark, protocol, or reward term, but \emph{Constraint State Governance}: constraints represented as execution state, carried across delegation, checked before critical actions, recorded for audit, and supplied to learning.

\begin{table*}[!htbp]
\centering
\caption{Coverage of mechanism families for LLM-agent safety. Existing mechanisms provide useful local controls, but they do not by themselves maintain constraints as inherited, checkable, and auditable state across full multi-agent trajectories.}
\label{tab:mechanism_limits}
\scriptsize
\setlength{\tabcolsep}{4pt}
\renewcommand{\arraystretch}{1.18}
\begin{tabularx}{\textwidth}{
>{\raggedright\arraybackslash}X
>{\centering\arraybackslash}p{0.075\textwidth}
>{\centering\arraybackslash}p{0.075\textwidth}
>{\centering\arraybackslash}p{0.075\textwidth}
>{\centering\arraybackslash}p{0.075\textwidth}
>{\centering\arraybackslash}p{0.075\textwidth}
>{\centering\arraybackslash}p{0.075\textwidth}
}
\toprule
\textbf{Mechanism family} &
\textbf{Local} &
\textbf{Tool} &
\textbf{Multi} &
\textbf{State} &
\textbf{Admit} &
\textbf{Audit} \\
\midrule
Response-level safeguards(prompts, refusals, filters) &
\cmark & \xmark & \xmark & \xmark & \pmark & \xmark \\

Tool and action safety(AgentDojo, InjecAgent, ToolEmu, WASP, OpenAgentSafety) &
\cmark & \cmark & \pmark & \xmark & \pmark & \pmark \\

Agent frameworks/protocols(MCP, A2A, LangGraph, AutoGen) &
\pmark & \cmark & \cmark & \pmark & \pmark & \xmark \\

Multi-agent safety benchmarks(AgentLeak, PAC-Bench, Colosseum) &
\pmark & \pmark & \cmark & \xmark & \xmark & \cmark \\

Delegation and capability systems(Capability Security, AIP, ACP) &
\pmark & \pmark & \cmark & \xmark & \cmark & \cmark \\

Safe RL and process supervision(CPO, RCPO, Process Supervision) &
\pmark & \xmark & \pmark & \xmark & \xmark & \pmark \\

\midrule
\textbf{Constraint State Governance} &
\cmark & \cmark & \cmark & \cmark$^\dagger$ & \cmark$^\dagger$ & \cmark$^\dagger$ \\
\bottomrule
\end{tabularx}

\vspace{0.35em}
\begin{flushleft}
\footnotesize
\textit{Notes.} \cmark = supported; \xmark = not supported; \pmark = partially supported; $\dagger$ = proposed design target. 
Local denotes local output or behavior control; Tool denotes tool or action-level control; Multi denotes multi-agent coordination; State denotes maintained constraint state; Admit denotes pre-action admission under active constraints; Audit denotes reconstructable evidence for why an action was allowed.
\end{flushleft}
\end{table*}

\section{Paradigm Design}

This section presents an operational blueprint for the position, not a complete cryptographic protocol or a complete reinforcement learning algorithm. The aim is to make the proposed paradigm specific enough for evaluation. The proposed paradigm has two coupled layers. Constraint State Governance (CSG) maintains the safety conditions that define whether a trajectory is permissible. Constraint Native Reinforcement Learning then improves utility only over trajectories whose constraints remain maintained. The point is not to attach RL after a governance module, but to make governance produce the very rollout substrate on which learning operates: signed state, scoped authority, admission decisions, realized actions, rejected actions, and audit evidence.

\subsection{Constraint State Governance}

Constraint drift occurs because a rule can be present in the initial prompt while losing force later in the trajectory: a downstream agent may act on a stale summary, a delegated task may silently expand into broader authority, sensitive context may cross an internal channel, or an unsafe action may be logged without enough evidence to explain why it was allowed. Constraint State Governance addresses this by making safety critical constraints explicit state objects rather than conversational reminders. The system does not rely on agents to remember a rule. It gives each rule an identity, version, scope, predicate, and audit trail.

A constraint is first converted into a signed token:
\begin{equation}
\rho_k =
\operatorname{Sign}_{G}
\big(
k,\ H(c_k),\ pred_k,\ scope_k,\ priority_k,\ ttl_k,\nu_k
\big).
\end{equation}
Here $c_k$ is the original rule text, $pred_k$ is the runtime predicate, $scope_k$ defines where the rule applies, $priority_k$ resolves conflicts, $ttl_k$ gives expiry, and $\nu_k$ is the rule version. This tokenization gives the rule a stable reference that can be inherited, checked, revoked, and audited. The hash binds the human readable rule to its executable predicate without repeatedly exposing the raw rule text.

At time $t$, the governance layer maintains:
\begin{equation}
B_t=\operatorname{Canon}(C_t,P_t,b_t,L_t,R_t,\nu_t),
\qquad
D_t=H(B_t\parallel r_t),
\qquad
\sigma_t=\operatorname{Sign}_{G}(D_t,t,\nu_t).
\end{equation}
$C_t$ is the active token set, $P_t$ the role and authority map, $b_t$ remaining budgets, $L_t$ labels or taints over sensitive objects, $R_t$ revocations, $\nu_t$ the state version, and $r_t$ the audit root. The signed digest $\sigma_t$ is the current state reference distributed to agents and tool wrappers. A critical action must therefore be checked against the current state digest, not against an agent's memory of what the task used to say. CSG builds on information flow control, taint tracking, capability security, authenticated delegation, admission control, and tamper evident logging \citep{10.1145/269005.266669, 10.1145/2619091, south2025authenticateddelegationauthorizedai, https://doi.org/10.5281/zenodo.19672575}. The algorithm is an illustrative runtime contract, not a complete semantic verifier. Its strongest form requires conservative effect extraction and sound rule predicates; weaker implementations can use channel policies, scoped tool wrappers, taint labels, and audit logs.

\begin{algorithm}[!htbp]
\caption{CSG admission for a governance critical action}
\label{alg:csg_admission}
\scriptsize
\begin{algorithmic}[1]
\Require State $B_t=(C_t,P_t,b_t,L_t,R_t,\nu_t)$, audit root $r_t$, signed state token $\sigma_t$
\Require Proposal $q=(i,a,\hat{\sigma},\kappa,\pi)$ from agent $i$
\State $D_t\leftarrow H(B_t\parallel r_t)$
\State $\phi\leftarrow \textsc{Effect}(a)$ 
\Comment{reads, writes, deletes, disclosures, labels, tool effects}
\State $ok\leftarrow \textsc{Fresh}(\hat{\sigma},D_t)$
\State $ok\leftarrow ok\land \textsc{CapOK}(\kappa,i,P_t,R_t)$
\Comment{$\kappa$ binds agent, task, scope, channels, budget, expiry, parent authority}
\State $ok\leftarrow ok\land \textsc{ScopeOK}(\phi,\kappa)$
\State $ok\leftarrow ok\land \textsc{FlowOK}(\phi,L_t,b_t)$
\State $ok\leftarrow ok\land \bigwedge_{\rho_k\in C_t} pred_k(\phi,\kappa,L_t,b_t)$
\State $ok\leftarrow ok\land \textsc{EvidenceOK}(\pi,\phi,C_t)$
\If{$ok$}
    \State execute $a$; $\chi\leftarrow 1$
\Else
    \State block $a$; $\chi\leftarrow 0$
\EndIf
\State $e_t\leftarrow\operatorname{Canon}(i,H(a),H(\kappa),H(\pi),D_t,\chi,\textit{reason},t)$
\State $r_{t+1}\leftarrow H(r_t\parallel H(e_t))$
\State \Return $\bar a=a$ if $\chi=1$, otherwise $\bar a=\bot$; updated root $r_{t+1}$
\end{algorithmic}
\vspace{-0.5em}
\end{algorithm}

The checks in Algorithm~\ref{alg:csg_admission} correspond directly to the drift modes. \textsc{Fresh} prevents stale state execution: a handoff that omits a newly added rule cannot authorize a critical action. \textsc{CapOK} and \textsc{ScopeOK} prevent authority drift: a delegated agent can act only within the capability it received. \textsc{FlowOK} prevents information flow drift: labels and budgets decide whether sensitive content can cross a channel before the transfer occurs. The event record prevents accountability drift: the system records the active digest, action hash, capability hash, evidence hash, decision, and reason at the time of admission, not after the agent has already acted.

\paragraph{Design invariant under idealized assumptions.}
Assume $H$ is collision resistant, signatures are unforgeable, every governance critical action is mediated by Algorithm~\ref{alg:csg_admission}, and each active token $\rho_k$ has a sound predicate over $\textsc{Effect}(a)$. If a critical action is realized, then it was admitted under the current signed constraint state, valid delegated authority, satisfied rule predicates, permitted information flow, and recorded audit evidence. After an audit root is announced, changing the recorded action, capability, evidence, decision, or reason without detection requires either a hash collision or a forged signature.

\emph{Proof sketch.}
A realized action requires $\chi=1$, so all checks in Algorithm~\ref{alg:csg_admission} passed. Freshness binds the proposal to the current digest $D_t$; capability validity binds the agent to scoped delegated authority; the scope, flow, rule, and evidence predicates enforce encoded constraints before execution; and the event $e_t$ commits to the action, capability, evidence, digest, decision, and reason. Since $r_{t+1}=H(r_t\parallel H(e_t))$, changing any recorded event changes the root unless the adversary finds a collision. Replacing the state token or capability requires a forged signature. In practice, \textsc{Effect} should be conservative: uncertain effects should trigger repair, sandboxing, or escalation rather than silent admission.

Consider the coding agent example. The user asks the system to fix an authentication bug while forbidding API key exposure, external disclosure of proprietary code, production configuration edits, test deletion, and unevidenced destructive commands. These rules become tokens in $C_t$; repository objects receive labels in $L_t$; the patching agent receives a capability for writes only under \texttt{src/auth/}; the debugging agent may run tests; and the cleanup agent may remove only temporary artifacts. A patch to \texttt{src/auth/login.py} with evidence from a failing authentication test can pass admission. An external query containing proprietary source code fails \textsc{FlowOK}. A command such as \texttt{rm tests/test\_auth.py} fails the test integrity or evidence predicate. The point is not that the model was reminded again. The action never enters the realized trajectory unless the signed state, scoped capability, rule predicates, and audit record agree.

\subsection{Constraint Native Reinforcement Learning}

CSG can block unsafe critical actions, but blocking alone does not make agents useful. The learning problem is not to add another safety penalty to a task reward. It is to train agents to find high utility trajectories inside a boundary maintained by signed constraint state, admission decisions, and audit events. This is where CSG fits naturally with reinforcement learning: the governance layer gives learning a verifiable rollout record, not just a textual instruction. Existing constrained RL, CMDPs, SafeMARL, CPO, RCPO, and safety oriented preference optimization already study how policies optimize under constraints \citep{achiam2017constrainedpolicyoptimization, tessler2018rewardconstrainedpolicyoptimization, kushwaha2026surveysafereinforcementlearning, lightman2023letsverifystepstep, shao2024deepseekmathpushinglimitsmathematical}. The difference here is earlier in the pipeline. Those methods usually assume that the constraint is available as a cost, label, preference, or observation. In LLM-based multi-agent systems, the constraint may have already drifted before learning sees the reward. Adding ``constraint state'' to the observation is not enough if delegation is stale, tool calls are unlogged, sensitive content moves through internal channels, or audit evidence is missing. Constraint native learning therefore uses CSG to define the comparison set itself.

For a trajectory initialized from signed constraint state $B_0$, write
\begin{equation}
\tau=(B_0,q_{1:T},\chi_{1:T},\bar a_{1:T},e_{1:T}),
\qquad
m(\tau)=
(m_{\mathrm{fresh}},m_{\mathrm{scope}},m_{\mathrm{flow}},m_{\mathrm{audit}}),
\end{equation}
where $q_{1:T}$ are proposed actions, $\chi_{1:T}$ are admission decisions, $\bar a_{1:T}$ are realized actions, $e_{1:T}$ are audit events, and $m(\tau)$ measures whether the trajectory preserves freshness, delegated scope, information flow, and auditability. The admissible set is
\begin{equation}
\mathcal{T}_{\mathrm{adm}}(B_0)
=
\{\tau: m_j(\tau)\ge \epsilon_j,\ \forall j\},
\qquad
\tau^\star
=
\arg\max_{\tau\in\mathcal{T}_{\mathrm{adm}}(B_0)} U(\tau).
\end{equation}
Thus mandatory constraints define the space in which utility is optimized; they are not terms that a better final answer can compensate for. For a group of sampled trajectories $\mathcal{G}(B_0)=\{\tau_1,\ldots,\tau_M\}$, let $\mathcal{G}_{\mathrm{adm}}=\mathcal{G}\cap\mathcal{T}_{\mathrm{adm}}(B_0)$. A compact learning signal is
\begin{equation}
A(\tau_m)
=
\mathbf{1}[\tau_m\in\mathcal{G}_{\mathrm{adm}}]\,
\frac{U(\tau_m)-\mu_{\mathrm{adm}}}{\sigma_{\mathrm{adm}}+\delta}
-
\lambda\,\mathbf{1}[\tau_m\notin\mathcal{G}_{\mathrm{adm}}]
\sum_j [\epsilon_j-m_j(\tau_m)]_+ ,
\end{equation}
where $\mu_{\mathrm{adm}}$ and $\sigma_{\mathrm{adm}}$ are computed only over admissible trajectories. The native advantage is the separation: admissible trajectories compete on usefulness, while inadmissible trajectories become violation evidence. A trajectory that breaks freshness, scope, information flow, or auditability is not a low quality success; it is outside the comparison set.

In the coding agent example, several trajectories may fix the same authentication bug. One edits \texttt{src/auth/login.py}, preserves tests, avoids external disclosure, and records evidence; it belongs to $\mathcal{T}_{\mathrm{adm}}(B_0)$ and can be ranked by patch quality, efficiency, and test success. Another sends proprietary source code to an external tool and obtains a better looking patch; it violates $m_{\mathrm{flow}}$. A third deletes \texttt{tests/test\_auth.py} and reports that all tests pass; it violates scope, audit, or test integrity. Ordinary task optimization may reward the latter two because they improve apparent success. Constraint native learning cannot: the signed state, admission trace, and audit events mark them as inadmissible before utility comparison. CSG maintains the boundary; RL improves behavior within it.

\subsection{Closed Loop Coupling}

The key point is that CSG does not sit outside learning as a safety wrapper. It changes the object that learning sees. A rollout is no longer just model outputs and rewards; it is a governed trajectory whose proposals, admission decisions, rejected actions, realized actions, and audit events are tied to signed state. Learning therefore receives not only task outcomes, but also a verifiable record of which constraints were preserved, which failed, and which actions remained admissible.

\begin{figure*}[!htbp]
\centering
\includegraphics[width=\textwidth]{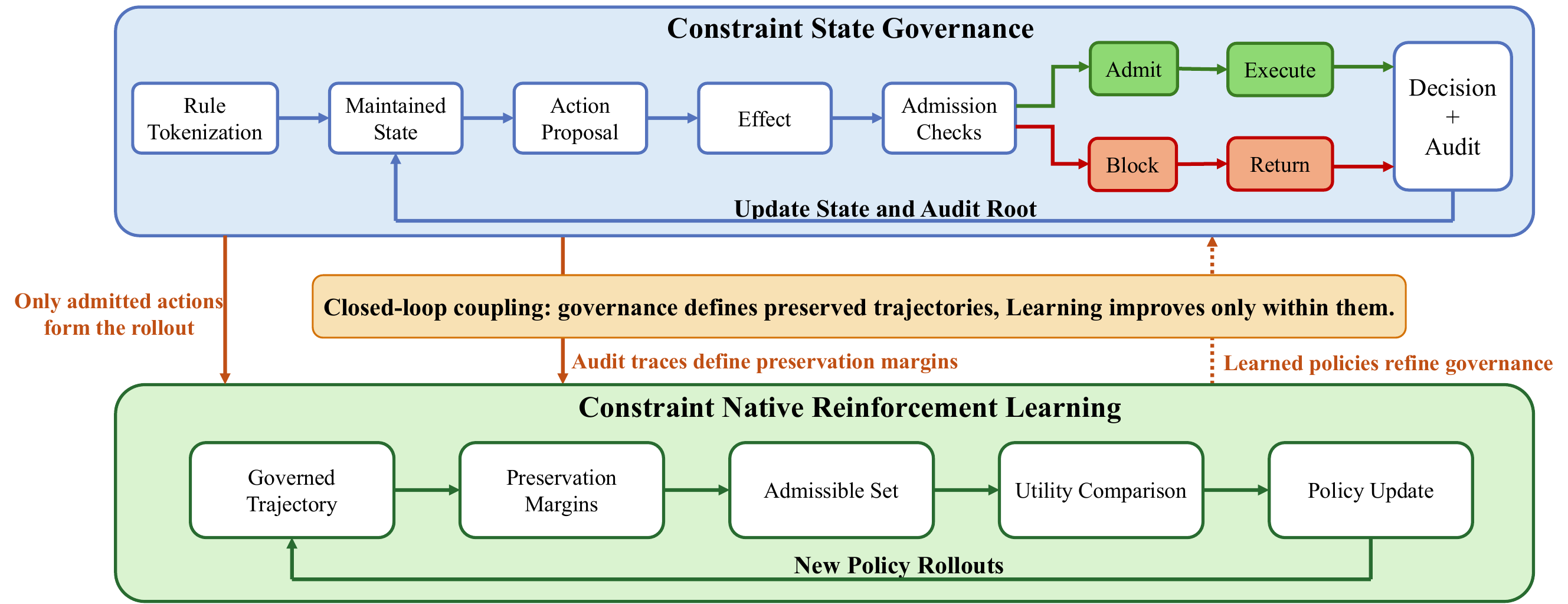}
\caption{CSG turns rules into governed trajectories, and constraint native learning improves utility only within the admissible set.}
\label{fig:csg_rl_workflow}
\end{figure*}

The closed loop can be summarized as:
\begin{equation}
(B_t,\pi_\theta)
\ \xrightarrow{\mathrm{CSG}}\ 
\tau=(q,\chi,\bar a,e)_{1:T}
\ \xrightarrow{\mathrm{score}}\ 
\big(m(\tau),U(\tau),v(\tau)\big)
\ \xrightarrow{\mathrm{learn}}\ 
(\pi_{\theta'},\Delta B).
\end{equation}
Here $\pi_\theta$ is the current policy, $m(\tau)$ are preservation margins, $U(\tau)$ is task utility, and $v(\tau)$ is violation evidence. The policy update $\pi_{\theta'}$ improves behavior inside the admissible set. The governance update $\Delta B$ is different: it is a proposed repair to predicates, scopes, budgets, labels, or escalation rules, and must itself pass through governance. Learning may reveal that a rule is vague, too strict, or often violated, but it cannot silently rewrite that rule for higher reward.

This coupling gives the paradigm its native advantage. A trajectory that leaks data or loses audit evidence cannot compete as a high utility success. At the same time, repeated violations can improve the governance layer by exposing missing labels, overbroad capabilities, brittle predicates, or cases that need escalation. Governance defines the admissible boundary; learning searches for useful behavior inside it; violation evidence improves both without allowing either to erase the other.

\section{Empirical Case Study}

To ground the argument, we use AgentLeak as a replay based case study of constraint drift in multi-agent LLM systems \citep{yagoubi2026agentleakfullstackbenchmarkprivacy}. The dataset contains $4{,}979$ traces from healthcare, finance, legal, and corporate domains, with three observed channels: final output (C1), inter agent messages (C2), and shared memory writes (C5). We instantiate \emph{CSG Lite} as a minimal replay implementation of Constraint State Governance: it checks each observed transition against an active disclosure policy, redacts forbidden sensitive content, and records channel, decision, reason, content hash, constraint digest, and event hash. This is counterfactual replay, not online deployment. CSG Lite instantiates only the information-flow and audit parts of CSG. It does not evaluate capability delegation, online replanning, or utility-induced drift. It asks which observed violations would have been visible and catchable if transition level admission had been present.

\begin{table}[!htbp]
\centering
\caption{Leakage rates under replay over $N{=}4{,}979$ AgentLeak traces.}
\label{tab:agentleak_csg_lite}
\small
\begin{tabular}{lcccccc}
\toprule
\textbf{Condition} & \textbf{C1} & \textbf{C2} & \textbf{C5} & \textbf{Internal} & \textbf{Total} & \textbf{Audit} \\
 & Output & Inter agent & Memory & C2$\cup$C5 & Exposure & Events \\
\midrule
No Defense & 27.2\% & 68.8\% & 46.7\% & 68.8\% & 68.9\% & 0 \\
Output Only Filter & 2.9\% & 68.8\% & 46.7\% & 68.8\% & 68.8\% & 1,556 \\
\textbf{CSG Lite} & \textbf{2.9\%} & \textbf{3.3\%} & \textbf{2.7\%} & \textbf{4.6\%} & \textbf{5.7\%} & \textbf{19,916} \\
\bottomrule
\end{tabular}
\end{table}

Table~\ref{tab:agentleak_csg_lite} shows why output level safety is insufficient. Output only filtering makes the visible answer look safe, but leaves internal exposure almost unchanged. CSG Lite reduces hidden exposure by moving the check to the channels where information actually moves, and turns final boundary monitoring into transition level audit evidence. This supports two concrete drift modes: information flow drift, because sensitive content crosses C2 or C5 before C1 is inspected; and accountability drift, because violations become reconstructable only when transition decisions are logged.

We then test whether the same governed traces can support learning rather than static filtering. We run offline policy search over the same traces. Each candidate policy decides what to redact on C1, C2, and C5. For each replayed trajectory:
\[
\begin{aligned}
\tau\in\mathcal{T}_{\mathrm{adm}}
&\iff
\mathrm{ForbiddenExposure}(\tau)\leq 2\%
\ \land\
\mathrm{AuditValid}(\tau)=1,\\
U(\tau)
&=
R_{\mathrm{allow}}
-0.3R_{\mathrm{over}}
-0.1R_{\mathrm{redact}} .
\end{aligned}
\]
$R_{\mathrm{allow}}$ measures retention of allowed final output fields, $R_{\mathrm{over}}$ measures over redaction of allowed fields, and $R_{\mathrm{redact}}$ measures total redaction cost. Policies are selected on a stratified train split and reported on a held out test split. All replayed policies produce valid hash linked audit chains; the comparison therefore focuses on admissibility and utility.

\begin{table}[!htbp]
\centering
\caption{Offline policy search on the held out test split. Admissible policies must keep forbidden exposure below 2\%.}
\label{tab:csg_policy_search}
\small
\begin{tabular}{lccccc}
\toprule
\textbf{Policy} & \textbf{Forb. Exp.}$\downarrow$ & \textbf{Adm. Rate}$\uparrow$ & \textbf{Retention}$\uparrow$ & \textbf{Over Red.}$\downarrow$ & \textbf{Utility}$\uparrow$ \\
\midrule
Output Only & 0.328 & 0.672 & 0.787 & 0.017 & 0.718 \\
Strict CSG & 0.000 & 1.000 & 0.000 & 0.081 & -0.392 \\
Field Aware CSG & 0.017 & 0.983 & 1.000 & 0.000 & 0.922 \\
Inter Agent Lenient & 0.017 & 0.983 & 1.000 & 0.000 & 0.967 \\
\textbf{Balanced CSG} & \textbf{0.017} & \textbf{0.983} & \textbf{1.000} & \textbf{0.000} & \textbf{0.980} \\
\bottomrule
\end{tabular}
\end{table}

The policy search gives a small feasibility check for constraint native learning. Output only control is not admissible because forbidden exposure remains high. Strict CSG is admissible but loses utility through over redaction. Balanced CSG stays inside the admissible set while preserving allowed information. This is not online RL, and it does not model how agents would adapt after redaction. It shows the substrate required by the paradigm is implementable: audit chains verify, admissible sets are computable, and utility can be optimized inside those sets rather than by trading constraints away.

\section{Alternative Views and Objections}

A first objection is that many safety critical constraints cannot be cleanly turned into state. User intentions are often vague, contextual, or incomplete. ``be careful with this data'' is not the same kind of object as an access control rule. CSG does not ask every human norm to become a formal predicate. It asks for an operational form only when a constraint decides whether a critical action is allowed. Some constraints can become scopes, budgets, labels, taints, permissions, or admission predicates. Others may stay tied to policy text, human review, or escalation. The aim is not perfect formalization. The aim is to avoid a system where file deletion, external disclosure, delegation, or memory writes ride on an agent's informal memory of an instruction.

A second objection is that constraints conflict or change during execution. Auditability may force the system to keep information that data minimization wants to drop. Efficiency may push against approval gates. Test integrity may push against emergency repair. CSG does not remove the need for hierarchy, priority rules, or institutional judgment. What it does is make the tradeoff explicit. A system should show which constraints are mandatory, which are defeasible, which need escalation, and which rule was the basis for admitting an action. That is better than letting an agent quietly trade privacy, authority, auditability, and utility inside an opaque trajectory.

A third concern is rigidity. If every action requires heavy governance, multi-agent systems may become slow or unusable. This is why CSG should govern critical transitions, not every token. Strong admission checks are most important for external disclosure, delegation, file deletion, production configuration changes, shell commands, memory writes, sensitive tool calls, and irreversible actions. Low risk local reasoning can use lighter checks. The design target is not maximal control. It is proportional control. Actions that can violate safety critical constraints should be checked before they touch the environment.

A fourth objection is that the framework may overclaim what governance and learning can guarantee. The abstraction of an action's effect can miss side effects. Labels and taints can be incomplete. Tools can be compromised. Adversarial users can find gaps. Inadmissible trajectories can hurt exploration. These are real limits. CSG is not a proof of end to end safety. Constraint native RL is not a replacement for sandboxing, least privilege, secure tool design, runtime monitoring, human oversight, or adversarial testing. The claim is narrower than that. Once a safety critical constraint is chosen and encoded, it should be kept as part of the execution substrate. Learning should improve utility inside that boundary, not quietly wear it down. This narrower claim is also what makes the proposal testable. We can measure whether constraints are inherited, checked, violated, repaired, and audited across a trajectory.

\section{Research Agenda}
The first priority is to build benchmarks that make constraint drift observable. A useful benchmark should contain multi step agent tasks with explicit safety critical constraints, full trajectory records, and failure cases where the final answer looks successful while the trajectory violates a constraint. The log should include agent messages, memory writes, tool calls, file edits, shell commands, delegation chains, rejected actions, and audit events. Evaluation should report task utility and constraint preservation separately: task success, cost, and patch quality on one side; leakage, stale state actions, unauthorized edits, test deletion, inheritance failure, and audit reconstructability on the other. The central question should not be only whether the agent solved the task, but which constraint failed, at which interface, and whether the failure was visible before the final output.

The second priority is reusable instrumentation for agent systems. Current frameworks make it easy to route tasks and call tools, but much harder to inspect the safety state of a trajectory. Future frameworks should expose standard records for proposed actions, semantic effects, active constraint state, delegated authority, admission decisions, rejection reasons, and audit roots. Without these records, researchers can only inspect final outputs or manually reconstruct failures. With them, constraint drift becomes a measurable systems property rather than an anecdotal failure mode.

The third priority is practical constraint compilation. Many constraints will never become perfect formal rules, but safety critical ones often have operational parts: scopes, labels, budgets, protected paths, forbidden channels, approval rules, revocations, and escalation triggers. Future work should study how to extract these from user instructions, tool schemas, repository metadata, organizational policy, and runtime context, and how to update them when constraints conflict or change. This is where CSG must become engineering practice rather than notation.

\section{Conclusion}

LLM-based multi-agent systems change the unit of safety: the question is no longer only whether the final answer is acceptable, but whether safety critical constraints remain operative while agents summarize context, delegate work, call tools, update memory, execute actions, and optimize for success. This paper identified the failure mode as \emph{constraint drift}, the loss of operational force of safety critical constraints across an agentic trajectory: a constraint may still appear in the initial prompt or final explanation, but no longer govern the action that matters. The proposed response is \emph{Constraint State Governance}, where constraints are maintained as signed state, inherited through delegation, checked before critical actions, recorded for audit, and supplied to learning; constraint native reinforcement learning then improves utility inside the admissible boundary rather than treating mandatory constraints as soft penalties. Our AgentLeak replay illustrates the stakes: output only filtering can make final answers look safe while internal channels remain exposed, whereas transition level governance makes those violations visible, auditable, and catchable. CSG does not solve every problem in agent safety or make every human norm formally complete, but it identifies a missing systems layer for LLM-based multi-agent infrastructure. Safe multi-agent behavior should not depend on agents remembering constraints more carefully; the system should make constraints harder to lose. Safe multi-agent behavior must be maintained, not merely asserted.

\bibliographystyle{plainnat}
\bibliography{references}

\end{document}